\documentclass[10pt,twocolumn,letterpaper]{article}

\usepackage{cvpr}              %
\usepackage[accsupp]{axessibility}  %

\usepackage{graphicx}
\usepackage{amsmath}
\usepackage{amssymb}
\usepackage{booktabs}
\usepackage{threeparttable}
\usepackage{tabularx,multirow}

\usepackage[pagebackref,breaklinks,colorlinks]{hyperref}

\usepackage[capitalize]{cleveref}
\crefname{section}{Sec.}{Secs.}
\Crefname{section}{Section}{Sections}
\Crefname{table}{Table}{Tables}
\crefname{table}{Tab.}{Tabs.}

\def\cvprPaperID{123} %
\def\confName{CVPR}
\def\confYear{2022}

\begin{document}

\title{Image Quality Assessment with Gradient Siamese Network}

\author{Heng Cong\textsuperscript{}\thanks{The authors have contributed equally} \qquad 
	Lingzhi Fu\textsuperscript{*} \qquad
	Rongyu Zhang\textsuperscript{*} \qquad
	Yusheng Zhang \\
	Hao Wang \qquad
	Jiarong He \qquad
	Jin Gao \\
	Interactive Entertainment Group of Netease Inc\\
	Guangzhou, China\\
	{\tt\small \{congheng, fulingzhi, zhangrongyu, zhangyusheng\}@corp.netease.com}\\
	{\tt\small \{wanghao42, gzhejiarong,  jgao\}@corp.netease.com}
}
\maketitle

\begin{abstract}
In this work, we introduce Gradient Siamese Network (GSN) for image quality assessment. The proposed method is skilled in capturing the gradient features between distorted images and reference images in full-reference image quality assessment (IQA) task. We utilize Central Differential Convolution to obtain both semantic features and detail difference hidden in image pair. Furthermore, spatial attention guides the network to concentrate on regions related to image detail. For the low-level, mid-level, and high-level features extracted by the network, we innovatively design a multi-level fusion method to improve the efficiency of feature utilization. In addition to the common mean square error supervision, we further consider the relative distance among batch samples and successfully apply KL divergence loss to the image quality assessment task. We experimented the proposed algorithm GSN on several publicly available datasets and proved its superior performance. Our network won the second place in NTIRE 2022 Perceptual Image Quality Assessment Challenge track 1 Full-Reference \cite{gu2022ntire}.
\end{abstract}

\section{Introduction}
\label{sec:intro}

Image quality assessment (IQA) plays an important role in most image-based tasks \cite{li2021infrared,li2022infrared,liu2022video}, which aims at replacing humans to accurately evaluate image quality. The existing IQA methods can be divided into objective and subjective quality evaluation. Human observers are in most cases the recipients of the image process task, so subjective image quality assessment (SIQA) is reliable. In SIQA, Mean opinion score (MOS) is currently the best image quality metric, where a large number of observers are required to evaluate image quality in a controlled testing environment \cite{streijl2016mean}. However, SIQA can provide accurate results but is slow and expensive in practice \cite{zhai2020perceptual}. By comparison, objective image quality assessment (OIQA) measures the image quality by using mathematical models, which are automatically and widely used \cite{narwaria2010objective,battisti2015objective,cadik2004evaluation,liu2011visual}. 

Existing OIQA methods have been successful, but are difficult in evaluating some specific image processing tasks \cite{jinjin2020pipal,ponomarenko2015image}. Traditional assessment metrics such as Peak Signal-to-Noise Ratio (PSNR) \cite{huynh2008scope} and structural similarity index (SSIM)\cite{wang2004image} are widely used to evaluate image quality, but they are ineffective for images obtained by GANs-based methods. Many improved SSIM-based methods\cite{wang2010information,flynn2013image,li2010content,chen2006gradient} can apply to different resolutions with a wider range of scenes, whereas the highly nonlinear characteristics of the human visual system are underutilized. In recent years, although deep learning techniques \cite{cheon2021perceptual,ayyoubzadeh2021asna,guo2021iqma,hammou2021egb,shi2021region} have achieved success in IQA tasks, various emerging image processing methods have brought new challenges to IQA methods. The method of extracting image gradient information by CNN is effective but ignores the key fine-grained information. Feature fusion-based IQA methods are used in many works but are inefficient. Traditional regression networks use an MSE function to supervise the network output, but do not take into account the logical relationship of image quality score ranking. To improve the IQA accuracy, we employ central differential convolution to gather low-quality image semantic features and detail differences, and image fine-grained is extracted. A multi-level feature fusion method is designed to use image feature information efficiently. In addition, we introduce KL scatter to further detect the relative scores of samples, and achieve better results in IQA tasks.

From the above, the critical problem of image quality assessment is the lack of robustness in extracting features in existing complex distortion scenes. Therefore, our work focuses on efficiently extracting contrast gradient information of image pairs in the full reference problem.  Our contributions can be summarized as follows:
\begin{itemize}
	\item We propose Gradient Siamese Network, namely GSN. The proposed siamese network built with central differential convolution (CDC) and spatial attention module are capable of extracting gradient and semantic features. Multi-scale feature fusion mechanism reconnects the features of different depths to mine for as much useful information as possible.
	\item We successfully apply KL divergence loss to image quality assessment task for raising the importance of relative order learning. 
	\item The proposed IQA method achieves 2nd place on the NTIRE 2022 IQA challenge track 1 full-reference.
\end{itemize}

The rest of this article is organized as follows. The following section presents the related work. Section 3 describes the proposed method and network architecture in detail. The experiments are introduced in Section 4. Finally, conclusions are given in Section 5.

\section{Related Work}
Previous image quality assessment based full-reference works are introduced in this section. We summary the Siamese network based IQA. The study of central difference convolution are also described.  

{\bf Image Quality Assessment.} According to the availability of the distortion-free reference image, objective IQA methods can be classified into three: full-reference (FR), reduce-reference (RR), and no-reference (NR) image quality evaluation. Compared with FR, the RR and NR methods reduce the dependence on the reference information, however, the image evaluation accuracy decreases with the absence of reference images. 
For FR methods, feature extraction is the key step. Initially, the traditional IQA methods such as mean square error (MSE), and PSNR are still used in image processing methods for IQA, which have low computational complexity. However, MSE and PSNR are more influenced by pixel points, less consistent with subjective evaluation, and ignore the human visual system (HVS). Considering HVS, Wang et al. \cite{wang2004image} proposed the SSIM by assuming that HVS is highly adapted to structural information. Following SSIM, multi-scale structural similarity index (MS-SSIM) \cite{wang2003multiscale} and information content weighted structural similarity index (IW-SSIM) \cite{wang2010information} are proposed to combine image details at different resolutions and viewing conditions for IQA. Larson argues that different strategies are used when HVS evaluates image quality, so the variations in local statistics of local brightness, contrast masking, and spatial frequency components are considered to find distortions in \cite{larson2010most}. By analyzing various angles of HVS, the FR-IQA algorithm performance and speed are improved. 
Deep-learning-based IQA methods have achieved success in various image processing tasks. Gao proposed DeepSim \cite{gao2017deepsim} to estimate the local similarities of features between distortion and reference images, which predict the final quality by pooling the local similarities. In \cite{bosse2017deep}, Bosse introduced a neural network to learn the local quality and local weights jointly, which allows for feature learning and regression in an end-to-end manner. In particular, the NTIRE 2021 Perceptual Image Quality Assessment Challenge \cite{Gu_2021_CVPR} has verified the effectiveness of deep-learning-based IQA methods. LIPT team\cite{cheon2021perceptual} is the winner of the NTIRE 2021 IQA challenge, which applies a transformer architecture to the perceptual IQA task. The second place in NTIRE 2021 IQA challenges is MT-GTD team. A new bilateral-branch multi-scale image quality estimation (IQMA) network\cite{guo2021iqma} is proposed by MT-GTD to extract multi-scale features from patches of the reference image and corresponding patches of the distorted image separately.

{\bf Siamese Network based IQA.} Siamese networks are used to measure the similarity of two inputs, in which two inputs are fed into two neural networks for forming a new feature representation. For the IQA task, an IQA method \cite{varga2020composition} is proposed that relies on a Siamese layout of pre-trained CNNs, which effectively learn the fine-grained, quality-aware features of images. Siamese network named LRSN \cite{niu2019siamese} is designed to rank the quality scores between the two image patches. LRSN uses Siamese CNN to achieve learning to rank for IQA. In NTIRE 2021 perceptual IQA challenge \cite{Gu_2021_CVPR}, the MACS team proposes three different networks based on a Siamese-Difference architecture\cite{ayyoubzadeh2021asna}, which is different from the traditional Siamese networks. By using Siamese-Difference architecture, the MACS team finished 9th in the tournament. The IQA network named RADN\cite{shi2021region} is proposed by THUIIGROUP1919, in which Siamese network architecture is used to combine a reference-oriented deformable convolution and patch-level attention module for the IQA task. THUIIGROUP1919 wins 4th place in the 2021 challenge. 

\begin{figure}[b]
	\centering
	\includegraphics[width=1\linewidth]{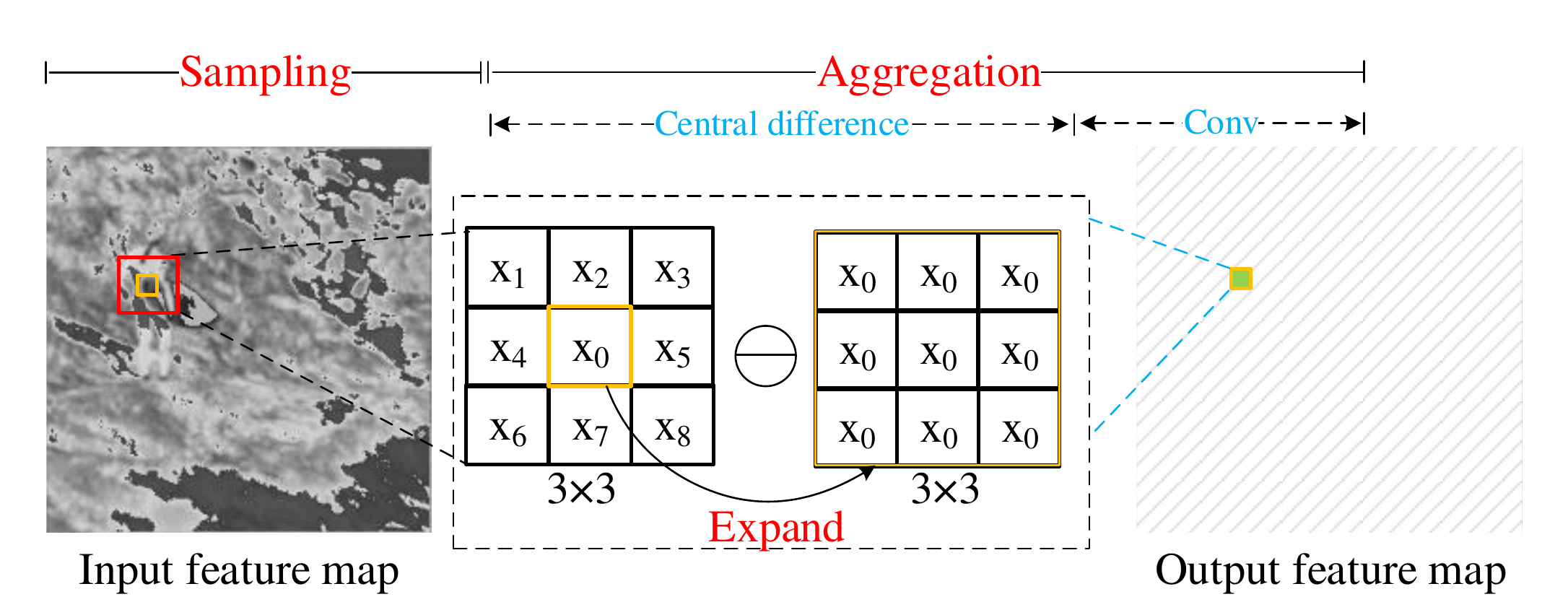}
	\caption{The central difference convolution.}
	\label{fig:CDCN}
\end{figure}

\begin{figure*}[t]
	\centering
	\includegraphics[width=1\linewidth]{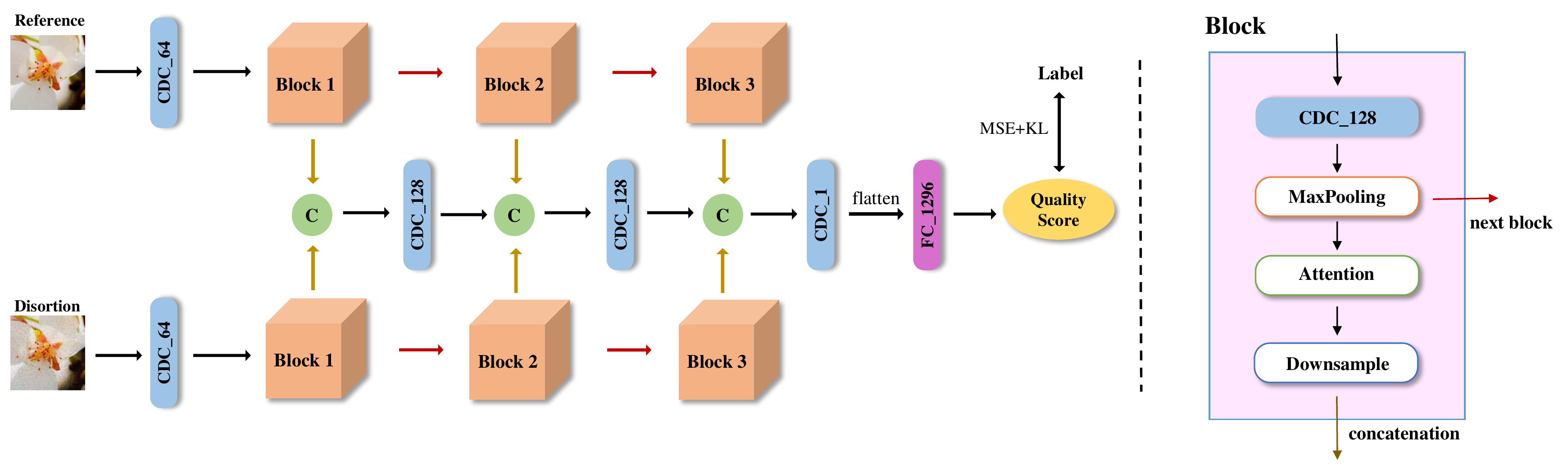}
	\caption{Model architecture of proposed gradient siamese network(GSN). C denotes concatenation and CDC\underline{ }64 indicates that the number of output channels is 64. After flattening the final features into a vector of length 1296, a fully connected layer marked as FC\underline{}1296 is introduced to obtain quality score. Note that the red line denotes passing onto next block and the golden line denotes passing onto multi-level feature fusion structure.}
	\label{fig:network}
\end{figure*}

{\bf Central Difference Convolution.} The central difference convolution (CDC) consists of sampling and aggregation, as shown in \cref{fig:CDCN}. Compared with the convolutional neural network (CNN), CDC extracts more fine-grained and robust features. Unlike the vanilla convolution \cite{xiao2018dynamical}, the aggression step in CDC focuses on the center-oriented gradient of sampled values. The intensity-level semantic information (ILSI) and gradient-level detailed message (GLDM) between the live and spoofing face in face anti-spoofing task \cite{yu2020searching} are balanced by CDC. This strategy could be implemented well in IQA tasks, given by ILSI and GLDM are crucial for distinguishing the distortion and reference image.

\section{Proposed Method}

The proposed GSN architecture is illustrated in \cref{fig:network}, which consists of a weight-sharing Siamese Network, a multi-level feature fusion module, and a regression head. First, we utilize a Siamese network based on CDC and spatial attention to process the reference image and distorted image. The network includes multi-level blocks, and the features extracted by each block are fixed to the same size by down-sampling. Secondly, we propose a multi-level feature fusion module that concatenates features at the same level and then fuses features from different levels. Finally, a fully connected layer is used as a regression head to obtain the predicted quality scores. 

\subsection{Siamese Network}
As shown in \cref{fig:network}, the basic block of the Siamese network includes Central Difference Convolution\cite{yu2020searching} and the spatial attention module\cite{woo2018cbam}. Central Difference Convolution, as the basic convolutional operation unit, can efficiently extract gradient features and semantic features for quality assessment. Then spatial attention module is used to obtain spatial attention features and improve the concentration on regions that are related to quality scores. The parameters settings of the network follow the study of neural architecture search in \cite{yu2020searching}.

Considering the greater similarity between the reference image and the distorted image in terms of image content, the gradient features are more significant and contribute more to the final score. This coincides with the face anti-spoofing task, so central difference convolution is introduced to obtain the gradient difference features between the reference image and the distorted image. CDC can replace the vanilla convolution, which can be simply implemented on the deep learning framework PyTorch\cite{paszke2017automatic} using the following equation:
\begin{equation}\label{key}
	\begin{aligned}
			y(p_{0})= \underbrace{\theta\cdot\sum_{p_{n}\in R} \omega(p_{n})\cdot(x(p_{0} + p_{n})-x(p_{0}))}_{central\, difference\, convolution} \\
			+ (1 - \theta)\cdot\underbrace{\sum_{p_{n}\in R} \omega(p_{n}) \cdot x(p_{0} + p_{n})}_{vanilla\, convolution},
	\end{aligned}
\end{equation}
where ${p_{0}}$ denotes the current position of the feature and ${p_{n}}$ denotes the other positions in the perceptual field ${R}$. The formula consists of central difference convolution and vanilla convolution, taking into account both intensity-level and gradient-level information. The hyper-parameter ${\theta}$ determines the weights of components. We set the value of ${\theta}$ as 0.7, indicating the importance and dependence on gradient information.

Spatial attention module considers the maximum peak and average values of the input features. After applying convolution, the output is constrained to [0, 1] with sigmoid function and then multiplied back to the input. The module calculation can be formulated as

\begin{equation}
	\begin{aligned}
		F_{i}^{'} = F_{i} \odot (\sigma C_{i}(A(F_{i}), M(F_{i}))),\\
		i\in\{low, mid, high\},
	\end{aligned}
\end{equation}
where $\odot$, ${A}$, and ${M}$ represent Hadamard product, average pooling, and max pooling, respectively. ${F_{low}}$, ${F_{mid}}$, and ${F_{high}}$ represent low-level feature, mid-level feature, and high-level feature. Especially, ${C}$ denotes vanilla convolution, where 7×7 convolution kernels are used in block 1 for low-level feature extraction. 5×5 kernels and 3×3 kernels are respectively used in block 2 (mid-level feature) and block 3 (high-level feature).

\begin{table*}[t]
	\centering
	\caption{GSN structure parameter settings.}
	\setlength{\tabcolsep}{2mm}{
	\begin{tabular}{cccccccc}
		\toprule
		& Layer & Size & Stride & Channel(input) & Channel(output) & BatchNorm & Activation \\
		\midrule
		conv1 & conv\_cdc & 3 & 1 & 3 & 64 & 64 & ReLU \\
		\midrule
		\multirow{5}*{block1} & conv\_cdc & 3 & 1 & 64 & 128 & 128 & ReLU \\
		& conv\_cdc & 3 & 1 & 128 & 204 & 204 & ReLU \\
		& conv\_cdc & 3 & 1 & 204 & 128 & 128 & ReLU \\
		& max-pooling & 3 & 2 & - & - & - & - \\
		& attention\_conv & 7 & 1 & 2 & 1 & - & Sigmoid \\
		\midrule
		\multirow{6}*{block2} & conv\_cdc & 3 & 1 & 128 & 153 & 153 & ReLU \\
		& conv\_cdc & 3 & 1 & 153 & 128 & 128 & ReLU \\
		& conv\_cdc & 3 & 1 & 128 & 179 & 179 & ReLU \\
		& conv\_cdc & 3 & 1 & 179 & 128 & 128 & ReLU \\
		& max-pooling & 3 & 2 & - & - & - & - \\
		& attention\_conv & 5 & 1 & 2 & 1 & - & Sigmoid \\
		\midrule
		\multirow{5}*{block3} & conv\_cdc & 3 & 1 & 128 & 128 & 128 & ReLU \\
		& conv\_cdc & 3 & 1 & 128 & 153 & 153 & ReLU \\
		& conv\_cdc & 3 & 1 & 153 & 128 & 128 & ReLU \\
		& max-pooling & 3 & 2 & - & - & - & - \\
		& attention\_conv & 3 & 1 & 2 & 1 & - & Sigmoid \\
		\bottomrule
	\end{tabular}
	}
	\label{tab:GSN Para}
\end{table*}

\subsection{Multi-level Feature Fusion}
We propose a multi-level feature fusion structure to reprocess features from different depths of the network, as shown in \cref{fig:network}. First, the lowest level features are concatenated together. Then after the number of channels is reduced by CDC, the features are gathered with mid-level features. Thus, the mid-level and high-level features are processed in turn. Finally, the feature dimension is compressed into one channel by CDC, flattened into 1296 dimensions and the final quality score is obtained with a fully connected layer.

\begin{table*}[t]
	\centering
	\caption{IQA datasets for comparison experiment.}
	\begin{tabular}{ccccccc}
		\toprule
		Database & No. Ref & No. Dist & No. Dist Type & No. Dist Level & Ground Truth & Resolution \\
		\midrule
		LIVE\cite{sheikh2006statistical} & 29 & 779 & 5 & 5 or 4 & MOS & $\sim$512×768\\
		CSIQ\cite{larson2010most} & 30 & 866 & 6 & 5 or 4 & MOS & 384×521\\
		TID2013\cite{ponomarenko2015image} & 25 & 3K &25 & 5 & MOS & 384×521\\
		KADID-10k\cite{lin2019kadid} & 81 & 10.1K & 25 & 5 & MOS & 512×512 \\
		NTIRE-2022 FR\cite{jinjin2020pipal} & 250 & 29K & 40 & 116 & MOS & 288×288\\ 
		\bottomrule
	\end{tabular}
	\label{tab:Dataset}
\end{table*}

\subsection{Architecture Details}
The parameter settings of GSN are shown in \cref{tab:GSN Para}. Given the RGB input, the first convolution layer expands the color channels to 64 dimensions and then features are fed into the three blocks in turn.

Note that output features of each block are fixed to the same size for multi-level feature fusion. Finally a fully connected layer transforms the feature vector into a quality score.

\subsection{Loss Function}
The novel loss function is designed to evaluate the predicted image scores and optimize the proposed model. The image quality assessment scores are normalized to make the predicted scores as close as possible to the subjective evaluation scores, given by

\begin{equation}
	\begin{aligned}
		\hat{S_{i}} = \dfrac{(\sum_{i=1}^{N}|\hat{Q_{i}} -\dfrac{1}{N} \sum_{i=1}^{N} \hat{Q_{i}}|^{q})^{\frac{1}{q}}}{\hat{Q_{i}} - \dfrac{1}{N} \sum_{i=1}^{N} \hat{Q_{i}}} (i=1,\cdots,N),
	\end{aligned}
\end{equation}

\begin{equation}
	\begin{aligned}
		S_{i} = \dfrac{(\sum_{i=1}^{N}|Q_{i} - \dfrac{1}{N} \sum_{i=1}^{N} Q_{i}|^{q})^{\frac{1}{q}}}{Q_{i} - \dfrac{1}{N} \sum_{i=1}^{N} Q_{i}} (i=1,\cdots,N),
	\end{aligned}
\end{equation}
where ${Q}$ and ${\hat{Q}}$ are the subjective evaluation scores and predicted quality scores, respectively. ${q\geq1}$ is a hyper-parameter, here, we set q as 2. %
${\hat{S}}$ and ${S}$ are the normalized predicted quality scores and the normalized subjected evaluation scores, respectively. The mean square error (MSE) is appiled and the function is defined as
\begin{equation}
	\begin{aligned}
		L_{pair}(Q, \hat{Q})=MSE(\hat{S_{i}}-S_{i}).
	\end{aligned}
\end{equation}
Additional, we apply Softmax regression for another set of normalized quality scores (${W}$ and  ${\hat{W}}$), given by

\begin{equation}
	\begin{aligned}
		\hat{W_{i}}=\dfrac{exp\,\hat{Q_{i}}}{\sum_{i=1}^{N} exp\,\hat{Q_{i}}} (i=1,\cdots,N),
	\end{aligned}
\end{equation}

\begin{equation}
	\begin{aligned}
		W_{i}=\dfrac{exp\,Q_{i}}{\sum_{i=1}^{N} exp\,Q_{i}} (i=1,\cdots,N).
	\end{aligned}
\end{equation}
The KL divergence between ${Q}$ and ${\hat{Q}}$ is calculated by
\begin{equation}
	\begin{aligned}
		L_{list} = D_{KL}(Q,\hat{Q}) = \sum_{i=1}^{N} \hat{W_{i}}\times log\dfrac{\hat{W_{i}}}{W_{i}}.
	\end{aligned}
\end{equation}
The total loss function can be defined as
\begin{equation}
	\begin{aligned}
		L(Q,\hat{Q}) = \alpha L_{pair}(Q,\hat{Q}) + \beta L_{list}(Q,\hat{Q}),
	\end{aligned}
\end{equation} 
where ${\alpha}$ and ${\beta}$ are both hyper-parameter. Here, ${\alpha=0.1}$ and ${\beta=0.1}$.

\section{Experiments}
The comparison experiments are implemented to demonstrate the effectiveness of our IQA method. Five public datasets are used to train and test for evaluating the proposed method. Ablation studies are conducted to investigate the behavior of GSN. The effectiveness, performance, and advantages of the proposed method are numerically and experimentally verified in this section.

\begin{table*}[t]
	\centering
	\begin{threeparttable}
		\caption{Comparison experiment of IQA methods on three standard IQA databases, i.e., LIVE \cite{sheikh2006statistical}, CSIQ \cite{larson2010most}, and TID2013 \cite{ponomarenko2015image}, in terms of PLCC, SRCC, KRCC and MS. The top two performing methods are highlighted in bold face. Some results are borrowed from \cite{ding2021comparison}.}
		\setlength{\tabcolsep}{2mm}{
				\begin{tabular}{ccccccccccccc}
					\toprule
					\multirow{2}{*}{Method}&
					\multicolumn{4}{c}{LIVE} &\multicolumn{4}{c}{CSIQ} &\multicolumn{4}{c}{TID2013}\cr
					\cmidrule(lr){2-5}\cmidrule(lr){6-9}\cmidrule(lr){10-13}
					
					&PLCC &SRCC & KRCC &MS &PLCC &SRCC & KRCC &MS &PLCC &SRCC & KRCC &MS\cr
					\midrule
					PSNR \cite{huynh2008scope} &0.865 &0.873 &0.68 &1.738 &0.819 &0.81 &0.601 &1.629 &0.677 &0.687 &0.496 &1.364 \cr
					SSIM \cite{wang2004image} & 0.937& 0.948 &0.796 &1.885 &0.852 &0.865 &0.68 &1.717 &0.777 &0.727 &0.545 &1.504\cr
					MS-SSIM \cite{wang2003multiscale} &0.940 &0.951 &0.805 &1.891 &0.889 &0.906 &0.730 &1.795 &0.830 &0.786 &0.605 &1.616\cr 
					VSI \cite{zhang2014vsi} &0.948 &0.952 &0.806 &1.900 &0.928 &0.942 &0.786 &1.870 &0.900 &0.897 &0.718 &{\bf1.797}\cr 
					MAD \cite{larson2010most} &0.968 &0.967 &0.842 &{\bf1.935} &0.95 &0.947 &0.797 &{\bf1.897} &0.827 &0.781 &0.604 &1.608\cr
					VIF \cite{sheikh2006image} &0.96 &0.964 &0.828 &1.924 &0.913 &0.911 &0.743 &1.824 &0.771 &0.677 &0.518 &1.448\cr
					FSIM \cite{zhang2011fsim} &0.961 &0.965 &0.836 &{\bf1.926} &0.919 &0.931 &0.769 &1.850 &0.877 &0.851 &0.667 &1.728\cr 
					NLPD \cite{laparra2016perceptual} &0.932 &0.937 &0.778 &1.869 &0.923 &0.932 &0.769 &1.855 &0.839 &0.800 &0.625 &1.639\cr
					GMSD \cite{xue2013gradient} &0.957 &0.96 &0.827 &1.917 &0.945 &0.950 &0.804 &1.895 &0.855 &0.804 &0.634 &1.659\cr
					\midrule
					DeepIQA \cite{bosse2017deep} &0.940 &0.947 &0.791 &1.887 &0.901 &0.909 &0.732 &1.810 &0.834 &0.831 &0.631 &1.665\cr
					PieAPP \cite{prashnani2018pieapp} &0.908 &0.919 &0.750 &1.827 &0.877 &0.892 &0.715 &1.769 &0.859 &0.876 &0.683 &1.735\cr
					LPIPS \cite{zhang2018unreasonable} &0.934 &0.932 &0.765 &1.866 &0.896 &0.876 &0.689 &1.772 &0.749 &0.67 &0.497 &1.419\cr
					DISTS \cite{ding2021comparison} &0.954 &0.954 &0.811 &1.908 &0.928 &0.929 &0.767 &1.857 &0.855 &0.830 &0.639 &1.685\cr
					\midrule
					GSN(ours) &0.922 &0.932 &0.764 &1.854 &0.944 &0.951 &0.801 &{\bf1.896} &0.892 &0.885 &0.699 &{\bf1.776}\cr
					\bottomrule
			\end{tabular}}
			\label{tab:compare}
		\end{threeparttable}
	\end{table*}

\begin{table}[t]
	\centering
	\begin{threeparttable}
		\caption{The results of comparion experiment on NTIRE 2022 \cite{gu2022ntire} FR Track. The top two performing methods are highlighted in bold face.}
		\setlength{\tabcolsep}{0.5mm}{
				\begin{tabular}{ccccc}
					\toprule
					\multirow{2}{*}{Method}&
					\multicolumn{2}{c}{Valid Dataset} &\multicolumn{2}{c}{Testing Dataset}\cr
					\cmidrule(lr){2-3}\cmidrule(lr){4-5}
					
					&MS &SRCC${/}$PLCC &MS &SRCC${/}$PLCC \cr
					\midrule
					${NTIRE_{2021}}$\cite{cheon2021perceptual}	& {\bf1.659} & {\bf0.819}${/}${\bf 0.840} &{\bf 1.588} & {\bf 0.789}${/}${\bf 0.798} \cr
					PSNR \cite{huynh2008scope}	&0.503 &0.233${/}$0.269 &0.526 &0.249${/}$0.276 \cr
					NQM \cite{damera2000image}	&0.666 &0.301${/}$0.364 &0.759 &0.364${/}$0.395 \cr
					UQI \cite{wang2002universal} &0.965 &0.460${/}$0.504 &0.869 &0.419${/}$0.450 \cr
					SSIM \cite{wang2004image} &0.696 &0.319${/}$0.377 &0.752 &0.361${/}$0.391 \cr
					MS-SSIM \cite{wang2003multiscale} &0.457 &0.337${/}$0.119 &0.532 &0.369${/}$0.163 \cr
					RFSIM \cite{zhang2010rfsim} &0.539 &0.253${/}$0.285 &0.632 &0.303${/}$0.328 \cr
					GSM \cite{liu2011image}	&0.828 &0.378${/}$0.450 &0.874 &0.409${/}$0.464 \cr
					SRSIM \cite{zhang2012sr} &1.155 &0.529${/}$0.626 &1.208 &0.572${/}$0.635 \cr
					FSIM \cite{zhang2011fsim} &1.005 &0.452${/}$0.552 &1.074 &0.503${/}$0.570 \cr
					VSI \cite{zhang2014vsi} &0.904 &0.411${/}$0.493 &0.975 &0.458${/}$0.516\cr
					NIQE \cite{mittal2012making} &0.141 &0.012${/}$0.128 &0.165 &0.034${/}$0.131\cr
					MA \cite{ma2017learning} &0.196 &0.099${/}$0.096 &0.287 &0.140${/}$0.146\cr
					PI \cite{blau2018perception} &0.198 &0.063${/}$0.134 &0.249 &0.103${/}$0.145\cr
					Brisque \cite{mittal2011blind} &0.059 &0.007${/}$0.051 &0.139 &0.070${/}$0.069\cr
					LPIPS-Alex \cite{zhang2018unreasonable} &1.174 &0.568${/}$0.606 &1.136 &0.565${/}$0.571\cr
					LPIPS-VGG \cite{zhang2018unreasonable} &1.161 &0.550${/}$0.610 &1.227 &0.594${/}$0.633\cr
					DISTS \cite{ding2021comparison}&1.242 &0.608${/}$0.634 &1.342 &0.654${/}$0.687\cr
					\midrule
					GSN (ours) &{\bf 1.684} &{\bf0.831}${/}${\bf0.853} &{\bf1.642} &{\bf0.815}${/}${\bf0.827} \cr
					\bottomrule
			\end{tabular}}
	\label{tab:NTIRE-2022}
		\end{threeparttable}
	\end{table}

\subsection{Dataset}
Comparison experiments are established based on three public databases, including the LIVE Image Quality Assessment Database (LIVE) \cite{sheikh2006statistical}, the Categorical Subjective Image Quality (CSIQ) database \cite{larson2010most}, and the Tampere image database 2013 (TID2013)\cite{ponomarenko2015image}. Furthermore, according to the official information provided, we evaluated the performance of the proposed method with other IQA algorithms on the NTIRE 2022 FR Track. With resolutions ranging from 438×634 pixels to 512×768 pixels, LIVE consisted of 29 reference images and 779 distorted images corrupted by 5 types of distortions. CSIQ includes 30 reference images and 866 distorted images corrupted by JPEG compression (JPEG), JPEG2000 compression (JP2K), white noise (WN), Gaussian blur (GB), additive pink Gaussian noise (AGB), and global contrast decrements. All images in CSIQ have a resolution of 384×521. TID2013 consists of 25 reference images and 3K distorted images, in which the distorted image is obtained by corrupting the reference image in 24 distorted types. For TID2013, all images have a resolution of 384×521. The NTIRE 2022 FR Track Dataset is derived from PIPAL\cite{jinjin2020pipal}, which includes 250 high-quality reference images (288×288 pixels), 40 distortion types, 29K distortion images, and 1130K human ratings. A lot of new distortion types, especially, a large number of image restoration algorithms’ results and GAN results are contained in PIPAL, which provides a more objective benchmark for the IQA model. In our comparison experiments, KADID-10k\cite{lin2019kadid} is used to train the proposed model, which has the most distortion types and distorted images. Specially, we use the PIPAL dataset for training and evaluation in NTIRE 2022 Perceptual IQA Challenge. PIPAL has 40 different distortion types, totally 29K distorted images. The above databases are detailed in \cref{tab:Dataset}.

\subsection{Implementation detail}

In the training phase, a given image pair is randomly horizontal flipped then cropped to obtain image patches fed into proposed GSN, which has dimension 192×192×3. In the testing phase, image patches are still acquired from the given image pair but sampled at fixed position. Extracting M overlapping patches (M = 5), we predict the final quality score by averaging respective quality scores of the M patches. Patches are distributed in the corners and center of the image.

The training is conducted using a Ranger optimizer with a batch size of 32. Initial learning rate is 1E-3 and weight decay is set as 1E-5. Epoch setting is 300. The training loss is computed using MSE loss function and KL Divergence loss function. GSN model is implemented using PyTorch framework(version $>=$ 1.4.0) and it roughly takes day and a half with a single Tesla V100 to train.

\subsection{Results}
After being trained on the entire KADID-10k dataset, our model was tested on three standard IQA databases LIVE, CSIQ and TID2013. Using Pearson linear correlation coefficient (PLCC), the Spearman rank correlation coefficient (SRCC), and the Kendall rank correlation coefficient (KRCC) as evaluation criteria, results are reported in \cref{tab:compare}. As same as the NTIRE 2022 Perceptual IQA Challenge \cite{gu2022ntire}, we list the sum of the PLCC and SRCC as main score criteria (MS) in the table. In addition, our model trained by the PIPAL \cite{jinjin2020pipal} is compared with \cite{cheon2021perceptual, huynh2008scope,damera2000image,wang2002universal, wang2004image,zhang2010rfsim,liu2011image,zhang2012sr,zhang2011fsim,zhang2014vsi,mittal2012making,ma2017learning,blau2018perception,mittal2011blind,zhang2018unreasonable,ding2021comparison} on NTIRE 2022 FR Track dataset for evaluating IQA performance. 

Experiments shown in \cref{tab:compare} demonstrate that our method has favorable performance compared to both classical methods (e.g., PSNR and SSIM) and recent deep-learning-based models (e.g., PieAPP, LPIPS, and DISTS). Specifically, the best and second-best main scores in LIVE are obtained by MAD and FSIM, respectively. The main reason is that the LIVE dataset is too small to measure the performance of IQA model well. The following experiments on CSIQ and TID2013 also show that the performance of MAD and FSIM will drop when they face a more diverse distortion type and level. For CSIQ and TID2013, our proposed GSN is more robust and ranked in the top two in terms of main score.

The performance comparison of IQA methods on NTIRE 2022 FR Track is shown in \cref{tab:NTIRE-2022}. It is worth noting that our method outperforms the NTIRE 2021 winner (${NTIRE_{2021}}$)\cite{cheon2021perceptual} in all metrics. Moreover, the best values in all evaluation criteria are achieved by our GSN, which proves the reliability of our method.

\begin{figure*}[htb]
	\centering
	\includegraphics[width=1\linewidth]{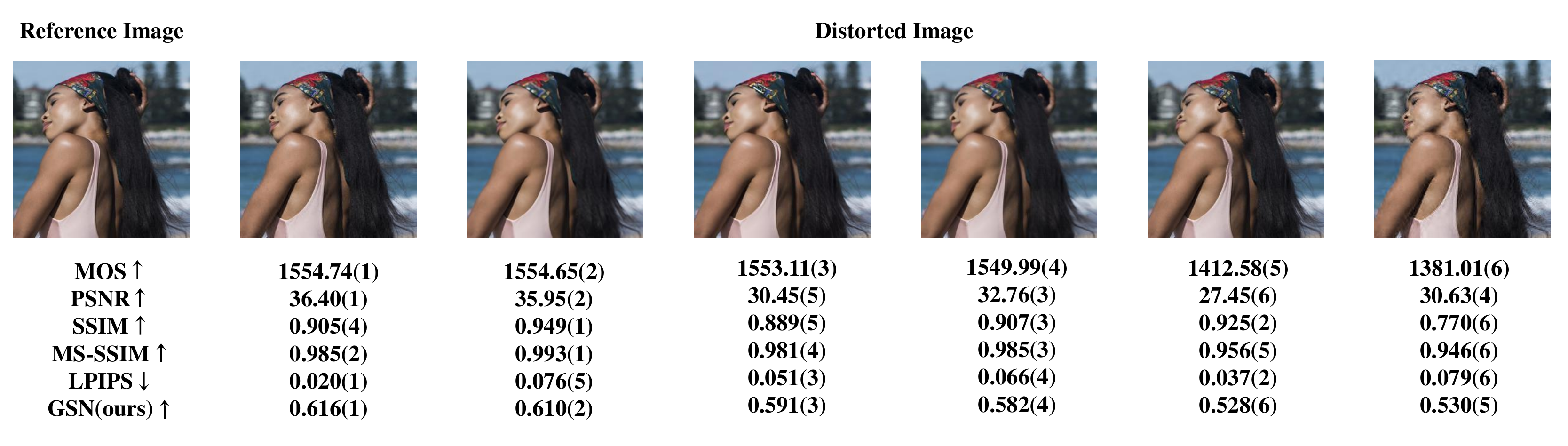}
	\caption{Visual result from the validation set of the NTIRE 2022 challenge. For each distortion, with MOS denoting the ground-truth human rating, predicted scores of PSNR \cite{huynh2008scope}, SSIM \cite{wang2004image}, MS-SSIM \cite{wang2003multiscale}, LPIPS \cite{zhang2018unreasonable}, and proposed GSN are listed. Up arrow indicates that the higher value the better quality while downward arrow indicates the opposite. The number in the parenthesis denotes the quality rank among distorted images in this figure.}
	\label{fig:sample}
\end{figure*}

\subsection{Ablations}
Extracting the image fine-grained and considering the relative ranking information are the key points to our GSN. In section 3, the CDC is used to obtain the image fine-grained, and KL divergence loss is proposed to improve prediction accuracy. To investigate the effect of convolution types and cost function, we design several ablation studies as shown in \cref{tab:Abtaion}. PIPAL \cite{jinjin2020pipal} dataset is used to test models 1-4 with the SRCC and PLCC as the evaluation metrics. Model 1 (M1) uses CNN and MSE as the convolution layer and loss function. The proposed loss function instead of MSE to supervise model 2 (M2) training. In contrast to M2, the CDC and MSE are used in model 3. Model 4 is our GSN, which includes CDC and the proposed function.

{\bf Effectiveness of CDC.} Comparing M1 (CNN${\And}$MSE) and M3 (CDC${\And}$MSE) in \cref{tab:Abtaion}, the PLCC and SRCC decreased by 0.043 and 0.053, respectively, after using CNN instead of CDC. M2 (CNN${\And}$MSE+KL) also differs from M4 (CDC${\And}$MSE+KL) in the choice of convolution type as shown in \cref{tab:Abtaion}, where the MS decreased by 0.057. These results suggest that extracting the image fine-grained is important for IQA tasks. All models using CDC can achieve an MS value over 1.6, which demonstrates that CDC is beneficial and effective in IQA.

{\bf Effectiveness of KL divergence.} By analyzing the relative ranking information, our model improves IQA accuracy. For PLCC and SRCC, the score earned with MSE+KL is higher than the one earned with MSE. Specifically, the MS score of M2 (CNN${\And}$MSE+KL) is 0.053 higher than the MS score of M1 (CNN${\And}$MSE) by adding KL function. Analyzing the MS score in M1 (CNN${\And}$MSE), M2 (CNN${\And}$MSE+KL), and M3 (CDC${\And}$MSE), although the M2 (CNN${\And}$MSE+KL) is lower 0.043 than M3 (CDC${\And}$MSE) owing to using CNN, M1 (CNN${\And}$MSE) is lower 0.096 than M3 (CDC${\And}$MSE). The results show that the induction of KL divergence can correct the score dropping and improve the predicted accuracy of IQA method.

\begin{table}[t]
	\centering
	\begin{threeparttable}
		\caption{The results of ablation studies. MS is the sum of PLCC and SRCC.}
		\setlength{\tabcolsep}{1mm}{
			\begin{tabular}{ccccccc}
				\toprule
				\multirow{2}{*}{Model}&
				\multicolumn{2}{c}{Conv.Type} &\multicolumn{2}{c}{Loss} & \multicolumn{2}{c}{PIPAL\cite{jinjin2020pipal}}\cr
				\cmidrule(lr){2-3}\cmidrule(lr){4-5}\cmidrule{6-7}
				
				&CDC &CNN &MSE &MSE+KL &PLCC${/}$SRCC &MS\cr
				\midrule
				M1 & &${\surd}$ &${\surd}$ & &0.785${/}$0.751 &1.536\cr
				M2 & &${\surd}$ & &${\surd}$ &0.802${/}$0.787 &1.589\cr
				M3 &${\surd}$ & &${\surd}$ & &0.828${/}$0.804 &1.632\cr
				M4 &${\surd}$ & & &${\surd}$ &{\bf0.835}${/}${\bf0.811} & {\bf1.646}\cr
				\bottomrule
		\end{tabular}}
		\label{tab:Abtaion}
	\end{threeparttable}
\end{table}

\begin{table}
	\centering
	\caption{Result of NTIRE 2022 IQA-FR Challenge.}
	\setlength{\tabcolsep}{4mm}{
	\begin{tabular}{c|c|c|c}
		Team & mainScore & PLCC & SRCC \\
		\hline
		1st & 1.651 & 0.828 & 0.822 \\
		\textbf{GSN(ours)} & \textbf{1.642} & \textbf{0.827} & \textbf{0.815} \\
		3rd & 1.640 & 0.823 & 0.817 \\
		4th  & 1.541 & 0.775 & 0.766\\
		5th  & 1.538 & 0.772 & 0.765\\
		6th  & 1.501 & 0.763 & 0.737\\
		7th  & 1.450 & 0.763 & 0.737\\
		8th  & 1.403 & 0.703 & 0.701\\
		\bottomrule
	\end{tabular}}
	\label{tab:NTIRE}
\end{table}

\begin{figure}[htb]
	\centering
	\includegraphics[width=1\linewidth]{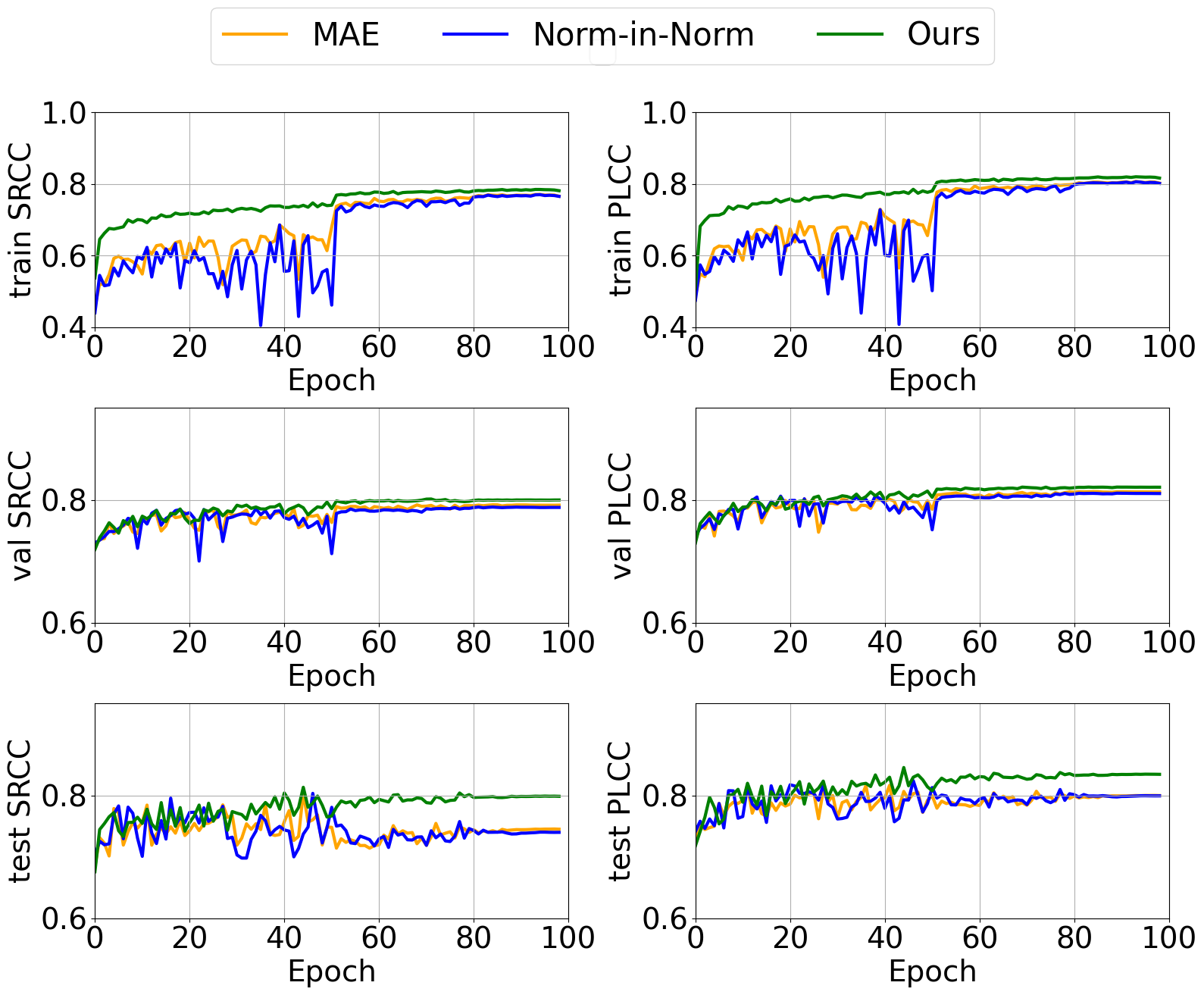}
	\caption{The training/validation/testing curves of the models trained with MAE, Norm-In-Norm, and our proposed loss. We train/validate the model on PIPAL dataset and test the model on TID2013 dataset.}
	\label{fig:curve}
\end{figure}

\begin{figure}[htb]
	\centering
	\includegraphics[width=1\linewidth]{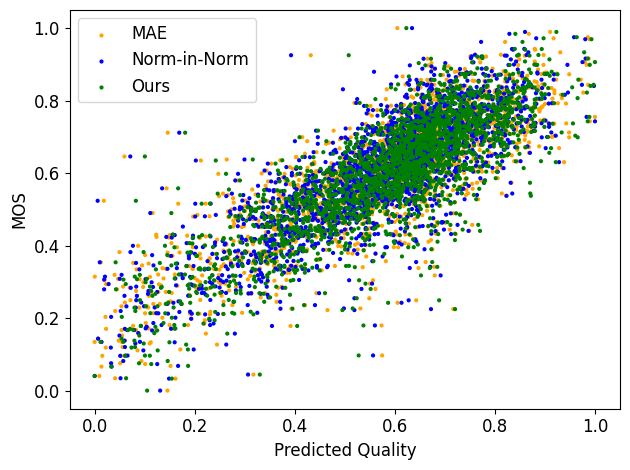}
	\caption{Scatter plot between the predicted quality and the MOS on PIPAL validation set.}
	\label{fig:scatter}
\end{figure}

\subsection{NTIRE 2022 Perceptual IQA Challenge}
We proposed our GSN to participate in the NTIRE 2022 Perceptual Image Quality Assessment Challenge \cite{gu2022ntire} Track 1 Full-Reference. The goal of this competition is to develop IQA methods with strong robustness and generalization capabilities that produce results showing high correlation with MOS scores. The NTIRE competition uses PIPAL as dataset. The PIPAL dataset is split into individual training sub-set, validation sub-set, and test sub-set, where the participants have MOS labels for the training set and no labels for the validation and test sets. The PIPAL dataset generated 23K distortion images by 40 distortion types and MOS labels by ELO rating system based on over 1 million human rankings. Both the validation set and the test set have 25 reference images and 1650 distortion images. This Challenge uses the main score as the scoring metric, which is a simple sum of PLCC and SRCC. The benchmark results of our model and the other teams in NTIRE 2022 are shown in \cref{tab:NTIRE} and the second place is won by us.

As shown as \cref{fig:sample}, the visual result of the PIPAL dataset predicted by PSNR, SSIM, MS-SSIM, LPIPS, and proposed GSN are demonstrated. The perceptually better to worse images are listed based on MOS from left to right. Our proposed GSN predictions are similar to ground-truth MOS, able to distinguish both distortion types with similar scores and local deformation situations. 

In ablations section, we have proven the effectivenss of KL divergence. The convergence results on PIPAL dataset and testing on TID2013 dataset are shown in \cref{fig:curve}. By using the proposed loss, the model achieved better prediction performance than MAE loss and Norm-In-Norm loss \cite{li2020norm}. The scatter plot between the predicted quality scores by the models and MOSs on PIPAL validation dataset set is shown in \cref{fig:scatter}. The scatter points of GSN with our designed loss are more centered in the diagonal line, which means a better prediction of image quality.

\section{Conclusion}
In this paper, we propose a novel network architecture, named gradient siamese network (GSN), for full-reference image quality assessment. Central difference convolution is introduced to GSN for the image fine-grained. To be consistent with the subjective human evaluation, residual features obtained by the proposed method are used to characterize the quality gap of the image pairs. A fusion structure is incorporated which handles multi-level features. With considering relative ranking information, we propose an innovative KL divergence loss to improve prediction accuracy. Testing statistics on several public IQA datasets demonstrate the superiority of the proposed GSN. Additionally, we ranked second in NTIRE 2022 Perceptual Image Quality Assessment Challenge Full-Reference.

{\small
\bibliographystyle{ieee_fullname}
\bibliography{egbib}
}

\end{document}